\documentclass[12pt]{article}
\newcommand{\Li}{\mathrm{Li}_2}
\newcommand{\DD}{{\mathcal D}}
\newcommand{\NS}{{\mathrm NS}}
\newcommand{\dd}{{\mathrm d}}

\def\fun#1#2{\lower3.6pt\vbox{\baselineskip0pt\lineskip.9pt
  \ialign{$\mathsurround=0pt#1\hfil##\hfil$\crcr#2\crcr\sim\crcr}}}

\title{Non--singlet splitting functions in QED}

\author{
A.B. Arbuzov\thanks{On leave of absence from 
Joint Institute for Nuclear Research, Dubna, Russia.}
}

\date{}

\begin{document}

\maketitle

\begin{itemize}
\item[$ $]
        {\em Dipartimento di Fisica Teorica, Universit\`a di Torino; \\
             INFN, Sezione di Torino, \\
             via Giuria 1, I-10125 Torino, Italy \/} \\
        {\tt e-mail: arbuzov@to.infn.it}
\end{itemize}

\begin{abstract}
Iterative solution of QED evolution equations for
non-singlet electron structure functions is considered.
Analytical expressions in the fourth and fifth orders
are presented in terms of splitting functions. Relation 
to the existing exponentiated solution is discussed.
\end{abstract}

\section{Introduction}

In this paper we are going to discuss properties of 
the QED non--singlet splitting functions. Some details of
derivation of the fourth and fifth order approximations are given.
The related subjects concerning the QED structure functions (SF)
themselves are touched very briefly, mainly to show, how do
the higher order splitting functions enter into the SF.
An extended discussion about the SF can be found in 
papers~\cite{Skrz,CDMN,KF,NT} and references therein. 
In Ref.~\cite{Prz} a fifth order
perturbative solution for the non--singlet SF was derived
within the {\it ad hoc} exponentiation procedure~\cite{JW}.

The Dokshitzer--Gribov--Lipatov--Altarelli--Parisi
evolution equation for the non--singlet
electron structure function reads
\begin{eqnarray} \label{eveq}
\DD^{\NS}(z,Q^2) =
\delta(1-z) + \int\limits_{m^2}^{Q^2}
\frac{\alpha(q^2)}{2\pi}\;\frac{\dd q^2}{q^2}
\int\limits_{z}^{1}\frac{\dd x}{x}\, P^{(1)}(x)
\DD^{\NS}(\frac{z}{x}, q^2), 
\end{eqnarray}
where $m$ is the electron mass; $P^{(1)}$ is the first order
non--singlet splitting function; 
$\alpha(q^2)$ is the QED running coupling
constant. Here we are going to consider only
the electron contribution to vacuum polarization:
\begin{eqnarray} \label{arun}
\alpha(q^2) = \frac{\alpha}{1-\frac{\alpha}{3\pi}
\ln\frac{q^2}{m^2}}.
\end{eqnarray}
It is worth noting that only the one--loop approximation 
(re--summed) gives the leading log contribution, while higher
orders provide only next--to--leading corrections.
The account of the running coupling constant 
in Eq.~(\ref{eveq})
is associated with the leading log radiative corrections  
due to pair production in the so--called non--singlet
channel. Usually pair production is considered separately
from the pure photonic correction, because of different
conditions of registration. Nevertheless, we are going
to evaluate the general equation, and then point out the
part, corresponding to pair production.
An account of other non--singlet pair contributions due to muons, 
$\tau$-leptons, and hadrons can be done within a certain
approximation in the final formulae.

The singlet pair production mechanism in the leading logarithmic 
approximation is described by the so--called singlet 
electron structure function $\DD^{\mathrm{S}}$ (see~\cite{Skrz,KF,NT});
we are not going to discuss it here.

\subsection{Splitting function}

Function $P^{(1)}(z)$ in Eq.~(\ref{eveq}) is the so--called
splitting function. It serves as a kernel function in the
evolution equation. The function (multiplied by the proper coefficient)
gives the first order 
leading log correction to the probability, 
which is just $\DD^{\NS}(z,Q^2)$, 
to find an electron (quark) with energy
fraction $z$ in the initial electron (quark): 
{\begin{eqnarray} \label{pee}
P^{(1)}(z) \equiv P_e^e(z) = \biggl(\frac{1+z^2}{1-z}\biggr)_+ 
= \frac{1+z^2}{1-z} 
- \delta(1-z)\int\limits_0^1\dd x\;\frac{1+x^2}{1-x}\, .
\end{eqnarray}
A simple derivation of the function can be found in Ref.~\cite{Alt}.

The splitting function is a generalised mathematical function. 
For practical applications we prefer to use the following 
definition, which allows to avoid explicitly the mutual
cancellation of infinite quantities during numerical 
computations: 
{\begin{eqnarray} \label{p1}
P^{(1)}(z) &=& \lim_{\Delta\to 0}\biggl\{ 2\biggl(\ln\Delta 
+ \frac{3}{4}\biggr)\delta(1-z) 
+ \frac{1+z^2}{1-z}\Theta(1-z-\Delta) \biggr\}
\nonumber \\
&\equiv& P^{(1)}_{\Delta}\delta(1-z) + P^{(1)}_{\Theta}\Theta(1-z-\Delta).
\end{eqnarray}
As could be seen from Eqs.~(\ref{pee},\ref{p1}), the function
satisfies the following normalisation condition:
{\begin{eqnarray} \label{norma}
\int\limits_0^1\dd z\; P^{(1)}(z) = 0.
\end{eqnarray}
This condition is a manifestation of the Kinoshita--Lee--Nauenberg 
theorem~\cite{KLN}: it provides the cancellation of mass singularities.

\section{Iterative solution}

In QED the evolution equation can be solved
to any desired order of perturbation theory by means of
iteration. The initial approximation for the structure
function is just $\delta(1-z)$. The first iteration 
gives 
\begin{eqnarray} \label{D1}
\DD^{\NS}(z,Q^2) = \delta(1-z)
+ \frac{\beta}{4}P^{(1)}(z) + {\mathcal O}(\alpha^2), \qquad
\beta = \frac{2\alpha}{\pi}(L-1), \qquad L=\ln\frac{Q^2}{m^2}\, ,
\end{eqnarray}
where $L$ is the so--called large logarithm. 

On the next step of the procedure we need to calculate the integral
\begin{eqnarray}
\int\limits_{z}^{1}\frac{\dd x}{x}P^{(1)}(x)P^{(1)}(\frac{z}{x})
\equiv P^{(1)}\otimes P^{(1)}(z).
\end{eqnarray}
This is a typical Mellin convolution
\begin{eqnarray} \label{conv}
P^{(n+1)}(z) = \int\limits_{z}^{1}\frac{\dd x}{x}P^{(1)}(x)
P^{(n)}(\frac{z}{x})
= \int\limits_{0}^{1}\dd x_1\int\limits^{1}_{0}\dd x_2\; 
P^{(1)}(x_1)P^{(n)}(x_2)\delta(z-x_1x_2).
\end{eqnarray}

In this way step by step we get the solution for the evolution equation
to the fifth order:
\begin{eqnarray}
\DD^{\NS}(z,Q^2) &=& \DD^{\NS}_{\gamma}(z,Q^2) + \DD^{\NS}_{e^+e^-}(z,Q^2),
\\  
\DD^{\NS}_{\gamma}(z,Q^2) &=& \delta(1-z) 
+ \sum_{n=1}^{5}\frac{1}{n!}\biggl(\frac{\beta}{4}\biggr)^nP^{(n)}(z)
+ {\mathcal O}(\alpha^6), \\
\DD^{\NS}_{e^+e^-}(z,Q^2) &=& \frac{1}{3}\biggl(\frac{\beta}{4}\biggr)^2
P^{(1)}(z) + \biggl(\frac{\beta}{4}\biggr)^3\biggl[ \frac{1}{3}P^{(2)}
+ \frac{4}{27}P^{(1)}\biggr] 
+ \biggl(\frac{\beta}{4}\biggr)^4\biggl[ \frac{1}{6}P^{(3)}
+ \frac{11}{54}P^{(2)} \nonumber \\
&+& \frac{2}{27}P^{(1)} \biggr] 
+ \biggl(\frac{\beta}{4}\biggr)^5\biggl[ \frac{1}{18}P^{(4)}
+ \frac{7}{54}P^{(3)} + \frac{10}{81}P^{(2)} + \frac{16}{405}P^{(1)}\biggr]
+ {\mathcal O}(\alpha^6). 
\end{eqnarray}
We denoted by index $\gamma$ the pure photonic part of the SF.
The other part describes pair corrections, and, starting
from the third order, with possible simultaneous photon radiation.
Recently we considered the numerical impact of the higher order 
pair corrections to electron--positron annihilation 
in paper~\cite{pairh}.

At the level of the non--singlet structure function 
the condition~(\ref{norma}) reads
\begin{eqnarray} \label{normad}
\int\limits_0^1\dd z\; \DD^{\NS}(z,Q^2) = 1, \qquad
\int\limits_0^1\dd z\; P^{(n)}(z) = 0, \quad n=1,2,\dots
\end{eqnarray}
This condition has also a trivial probabilistic meaning:
the sum of probabilities of all allowed emission processes
is unit. In other words, one is always able to find an electron
in the initial electron. 

Now we return to the splitting function properties.
Using prescription~(\ref{p1}) we can represent the integral of
the product of our two generalised functions in the form
\begin{eqnarray} \label{presc}
P^{(n+1)}(z) &=& P^{(n+1)}_{\Delta}\delta(1-z) 
+ P^{(n+1)}_{\Theta}\Theta(1-z-\Delta), \qquad \Delta\to 0,
\nonumber \\
P^{(n+1)}_{\Theta}(z) &=& P^{(1)}_{\Theta}(z)P^{(n)}_{\Delta} 
+ P^{(1)}_{\Delta}P^{(n)}_{\Theta}(z) 
+ \int\limits_{z/(1-\Delta)}^{1-\Delta}\frac{\dd x}{x}
P^{(1)}_{\Theta}(x)P^{(n)}_{\Theta}(\frac{z}{x}).
\end{eqnarray}
The $\Delta$-part of the splitting function can be obtained from the
condition~(\ref{norma},\ref{normad}):
\begin{eqnarray}
P^{(n+1)}_{\Delta} = - \int\limits_1^{1-\Delta}\dd z\;
P^{(n+1)}_{\Theta}(z).  
\end{eqnarray}
Instead of the above trick we can use, as was discussed in~\cite{Skrz},
the known solution of the
evolution equation in the soft limit~\cite{GL}:
\begin{eqnarray} \label{Gribov}
\DD^{\NS}_{\gamma}(z,Q^2)\bigg|_{z\to 1} = \frac{\beta}{2}\;
\frac{(1-z)^{\beta/2-1}}{\Gamma(1+\beta/2)}
\exp\biggl\{ \frac{\beta}{2}\biggl(\frac{3}{4}-C\biggr)
\biggr\},
\end{eqnarray}
where $C$ is the Euler constant $(C\approx 0.57721566)$. 
In order to obtain the $\Delta$-part
of a splitting function of the desired order we have to integrate 
Eq.~(\ref{Gribov}) over the interval $1-\Delta<z<1$, and then expand into
a series in $\alpha$:
\begin{eqnarray} 
\int\limits_{1-\Delta}^{1}\dd z\;\DD^{\NS}_{\gamma}(z,Q^2)
= \exp\biggl\{\frac{\beta}{2}\ln\Delta+\frac{3\beta}{8}\biggr\}
\frac{\exp(-C\beta/2)}{\Gamma(1+\beta/2)}\, .
\end{eqnarray}
Now we have to expand the exponent and use also the following formula
(see Appendix):
\begin{eqnarray} \label{expand}
\frac{\exp(-C\beta/2)}{\Gamma(1+\beta/2)} &=&
1 - \frac{1}{2}\biggl(\frac{\beta}{2}\biggr)^2\zeta(2) 
+ \frac{1}{3}\biggl(\frac{\beta}{2}\biggr)^3\zeta(3) 
+ \frac{1}{16}\biggl(\frac{\beta}{2}\biggr)^4\zeta(4) \nonumber \\
&+& \frac{1}{5}\biggl(\frac{\beta}{2}\biggr)^5\zeta(5)
- \frac{1}{6}\biggl(\frac{\beta}{2}\biggr)^5\zeta(2)\zeta(3)
+ {\mathcal O}(\beta^6). 
\end{eqnarray}

The second and third order splitting functions are well known 
(see paper~\cite{Skrz} and references therein) and used
in many various applications. For the sake of completeness
we put here the expressions:
\begin{eqnarray}
P^{(2)}_{\Theta}(z) &=& 2\biggl[ \frac{1+z^2}{1-z}\biggl(
2\ln(1-z) - \ln z + \frac{3}{2} \biggr) + \frac{1+z}{2}\ln z - 1 + z \biggr],
\nonumber \\
P^{(2)}_{\Delta} &=& 4\biggl(\ln\Delta + \frac{3}{4} \biggr)^2 - 4\zeta(2),
\\ \nonumber 
P^{(3)}_{\Theta}(z) &=& 24\frac{1+z^2}{1-z}\biggl( \frac{1}{2}\ln^2(1-z)
+ \frac{3}{4}\ln(1-z) - \frac{1}{2}\ln{z}\ln(1-z)
+ \frac{1}{12}\ln^2z \\ \nonumber
&-& \frac{3}{8}\ln{z} + \frac{9}{32} - \frac{1}{2}\zeta(2) \biggr)  
+ 6(1+z)\ln{z}\ln(1-z) - 12(1-z)\ln(1-z)  \\ \nonumber
&+& \frac{3}{2}(5-3z)\ln{z}
- 3(1-z) - \frac{3}{2}(1+z)\ln^2z + 6(1+z)\Li(1-z),
\nonumber \\
P^{(3)}_{\Delta} &=& 8\biggl(\ln\Delta + \frac{3}{4} \biggr)^3
- 24\zeta(2)\biggl(\ln\Delta + \frac{3}{4} \biggr) + 16\zeta(3).
\end{eqnarray}
The definition of the Riemann $\zeta$--functions, dilogarithm, 
and other special functions are given in Appendix. 

By means of the convolution procedure~(\ref{conv},\ref{presc}) we found
\begin{eqnarray}
P^{(4)}_{\Theta}(z) &=& 144\biggl\{ \frac{1+z^2}{1-z}\biggl[
\frac{2}{9}\ln^3(1-z) + \frac{1}{2}\ln^2(1-z) + \biggl( \frac{3}{8}
- \frac{2}{3}\zeta(2) \biggr)\ln(1-z)  
  \nonumber \\
&-& \frac{1}{3}\ln^2(1-z)\ln z
+ \frac{1}{9}\ln(1-z)\ln^2z - \frac{1}{2}\ln(1-z)\ln z 
- \frac{\ln^3z}{108} + \frac{\ln^2z}{12} 
\nonumber \\ 
&+& \biggl( \frac{\zeta(2)}{3} - \frac{3}{16} \biggr)\ln z
- \frac{1}{9}\ln z\Li(1-z)  - \frac{2}{9}{\mathrm{S}}_{1,2}(1-z)
+ \frac{3}{32} - \frac{\zeta(2)}{2} + \frac{4}{9}\zeta(3) \biggr]
\nonumber \\  
&-& \frac{1-z}{3}\ln^2(1-z) - \frac{1-z}{6}\ln(1-z)
+ \frac{1+z}{6}\ln^2(1-z)\ln z 
\nonumber \\     
&-& \frac{1+z}{12}\ln(1-z)\ln^2z
+ \frac{5-3z}{12}\ln(1-z)\ln z + \frac{7(1+z)}{864}\ln^3z
+ \frac{5z-11}{144}\ln^2z 
\nonumber \\     
&+& \biggl( \frac{43-5z}{288} 
- \frac{1+z}{6}\zeta(2) \biggr)\ln z 
+ \frac{1+z}{3}\ln(1-z)\Li(1-z) - \frac{1+z}{3}{\mathrm{Li}}_3(1-z)
\nonumber \\     
&+& \frac{1+z}{6}{\mathrm{S}}_{1,2}(1-z)
+ \frac{1+z}{12}{\mathrm{Li}}_2(1-z)
- \frac{11(1-z)}{144} + \frac{1-z}{3}\zeta(2)
\biggr\},
\end{eqnarray}

\begin{eqnarray}
P^{(4)}_{\Delta} &=& 16\biggl(\ln\Delta+\frac{3}{4}\biggr)^4
- 96\zeta(2)\biggl(\ln\Delta+\frac{3}{4}\biggr)^2
+ 128\zeta(3)\biggl(\ln\Delta+\frac{3}{4}\biggr)
+ 24\zeta(4).
\end{eqnarray}

At the next step we got
\begin{eqnarray}
P^{(5)}_{\Theta}(z) &=& 720 \Biggl\{ \frac{1+z^2}{1-z}\Biggl[
\frac{1}{9}\ln^4(1-z) + \frac{1}{3}\ln^3(1-z) 
+ \biggl( \frac{3}{8} - \frac{2}{3}\zeta(2) \biggr)\ln^2(1-z)
+ \biggl( \frac{3}{16} 
\nonumber \\
&-& \zeta(2) + \frac{8}{9}\zeta(3) \biggr)\ln(1-z) 
- \frac{2}{9}\ln{z}\ln^3(1-z) + \biggl( \frac{1}{9}\ln^2{z}
- \frac{1}{2}\ln{z}\biggr)\ln^2(1-z)
\nonumber \\
&+& \biggl( - \frac{1}{54}\ln^3z
+ \frac{1}{6}\ln^2z
- \biggl(\frac{2}{9}\Li(1-z) + \frac{3}{8}
- \frac{2}{3}\zeta(2)\biggr)\ln{z} 
- \frac{4}{9}{\mathrm{S}}_{1,2}(1-z)
\biggr)\ln(1-z)
\nonumber \\
&+& \frac{1}{1080}\ln^4z - \frac{1}{72}\ln^3z 
+ \biggl(\frac{1}{16} - \frac{1}{9}\zeta(2)
+ \frac{1}{18}\Li(1-z)
\biggr)\ln^2z
+ \biggl( - \frac{3}{32} 
+ \frac{1}{2}\zeta(2) 
\nonumber \\
&-& \frac{4}{9}\zeta(3)
+ \frac{1}{9}{\mathrm{S}}_{1,2}(1-z) - \frac{1}{6}\Li(1-z)
+ \frac{2}{9}{\mathrm{Li}}_{3}(1-z) \biggr)\ln{z}
+ \frac{1}{9}\Li^2(1-z) 
\nonumber \\
&-& \frac{1}{3}{\mathrm{S}}_{1,2}(1-z)
+ \frac{9}{256} - \frac{3}{8}\zeta(2) 
+ \frac{2}{3}\zeta(3) + \frac{1}{6}\zeta(4) \Biggr]
+ \biggl(\frac{(1+z)}{9}\ln{z} 
\nonumber \\
&-& \frac{2(1-z)}{9}\biggr)\ln^3(1-z) 
+ \biggl( - \frac{(1+z)}{12}\ln^2{z} 
+ \frac{5-3z}{12}\ln{z} + \frac{(1+z)}{3}\Li(1-z) 
\nonumber \\
&-& \frac{1-z}{6}\biggr)\ln^2(1-z) 
+ \biggl(\frac{7(1+z)}{432}\ln^3z + \frac{5z-11}{72}\ln^2z
+ \frac{43-5z}{144}\ln{z} 
\nonumber \\
&-& \frac{(1+z)}{3}\zeta(2)\ln{z}
+ \frac{(1+z)}{6}\Li(1-z)
+ \frac{(1+z)}{3}{\mathrm{S}}_{1,2}(1-z)
- \frac{2(1+z)}{3}{\mathrm{Li}}_{3}(1-z)
\nonumber \\
&+& \frac{2(1-z)}{3}\zeta(2) 
- \frac{11(1-z)}{72} \biggr)\ln(1-z)
- \frac{(1+z)}{1152}\ln^4z
+ \frac{23-9z}{1728}\ln^3z
\nonumber \\
&+& \biggl( \frac{(1+z)}{12}\zeta(2) 
- \frac{33+5z}{576} 
- \frac{5(1+z)}{144}\Li(1-z)\biggr)\ln^2z
+ \biggl(\frac{(1-z)}{9}\Li(1-z)
\nonumber \\
&-& \frac{(1+z)}{8}{\mathrm{S}}_{1,2}(1-z)
+ \frac{49-39z}{576} + \frac{2(1+z)}{9}\zeta(3)
- \frac{5-3z}{12} \biggr)\ln{z}
+ \biggl( \frac{19(1+z)}{144} 
\nonumber \\
&-&\frac{(1+z)}{3}\zeta(2) \biggr)\Li(1-z)
+ \frac{11-5z}{36}{\mathrm{S}}_{1,2}(1-z)
- \frac{(1+z)}{6}{\mathrm{Li}}_{3}(1-z)
\nonumber \\
&+& \frac{2(1+z)}{3}{\mathrm{Li}}_{4}(1-z)
- \frac{(1+z)}{3}{\mathrm{S}}_{2,2}(1-z)
- \frac{5(1+z)}{72}{\mathrm{S}}_{1,3}(1-z)
\nonumber \\
&-& \frac{5(1-z)}{288} + \frac{(1-z)}{6}\zeta(2)
- \frac{4(1-z)}{9}\zeta(3)
\Biggr\},
\end{eqnarray}

\begin{eqnarray}
P^{(5)}_{\Delta} &=& 32\biggl(\ln\Delta+\frac{3}{4}\biggr)^5
- 320\zeta(2)\biggl(\ln\Delta+\frac{3}{4}\biggr)^3
+ 640\zeta(3)\biggl(\ln\Delta+\frac{3}{4}\biggr)^2 \nonumber \\
&+& 240\zeta(4)\biggl(\ln\Delta+\frac{3}{4}\biggr)
+ 768\zeta(5) - 640\zeta(2)\zeta(3).
\end{eqnarray}

\section{Conclusions}

\begin{table}[ht]
\caption{Integral $I(x)$ in different approximations.}
\begin{tabular}[]{|l|c|c|c|c|c|c|} \hline
$x$ & ${\mathcal O}(\alpha)$ & ${\mathcal O}(\alpha^2)$ & 
${\mathcal O}(\alpha^3)$ & ${\mathcal O}(\alpha^4)$ & 
${\mathcal O}(\alpha^5)$ & exponent. \\ \hline
0.01 & 0.99972722 & 0.99970278 & 0.99970216 & 0.99970216 & 0.99970216 &
0.99970216 \\
0.1  & 0.99713067 & 0.99696386 & 0.99696237 & 0.99696241 & 0.99696241 &
0.99696241 \\
0.5  & 0.97933805 & 0.97871205 & 0.97873132 & 0.97873149 & 0.97873148 &
0.97873148 \\
0.9  & 0.91043156 & 0.91243286 & 0.91253629 & 0.91252777 & 0.91252803 &
0.91252802 \\
0.99 & 0.79019566 & 0.80982514 & 0.80884689 & 0.80886291 & 0.80886436 &
0.80886422 \\
0.999& 0.66569598 & 0.71915664 & 0.71380222 & 0.71416672 & 0.71415030 &
0.71415065 \\ \hline
\end{tabular}
\end{table}

The expressions obtained for the fourth and fifth order splitting
functions were checked to satisfy the condition~(\ref{normad}). In
this way we see the agreement between the iteration procedure 
for $P^{(n)}_{\Theta}(z)$ and the expansion of the known solution
for $P^{(n)}_{\Delta}$. Our result does also coincide with the
corresponding expansion of the exponentiated solution from 
Ref.~\cite{Prz}. So, we reproduced the known result, but in 
a different approach; and the higher order splitting functions
are given explicitly. The exponentiation and 
order--by--order calculations are complementary to each other.
As could be seen from Table~1, the numerical difference between
the exponentiated and non--exponentiated results is negligible,
and one may choose safely the approach, which he likes.
The results of our paper can be used to estimate higher order
radiative correction and to analyse the numerical difference
between exponentiated and order--by--order calculations.
Here we should note, that in a realistic situation 
in order to provide a high theoretical precision one should
take into account also sub--leading radiative corrections, 
which can be obtained only by direct perturbative calculations.

In Table~1 the values of integral 
\begin{eqnarray} 
I(x) = \int\limits_{x}^{1}\dd z\; \DD^{\NS}_{\gamma}(z,Q^2)
\end{eqnarray}
of the pure photonic part of the non--singlet SF is given for different
order approximations for $Q^2 = 10^4$~GeV$^2$, $L \approx 24.37$. 
For the last column the exponentiated result 
is obtained by using formula~(11) from Ref.~\cite{Prz}.
In Table~2 we present the corresponding values of integrals of the
splitting functions themselves:
\begin{eqnarray} 
J^{(n)}(x) = \int\limits_{x}^{1}\dd z\; P^{(n)}(z).
\end{eqnarray}
In the last line of Table~2 we put also the values of the corresponding
$\Delta$-parts. Note, that in reality, to simulate the limit $\Delta\to 0$,
and so to eliminate the dependence on $\Delta$ of numbers for integrals
$I(x)$ and $J^{(n)}(x)$, we used $\Delta=10^{-10}$. 
\begin{table}[ht]
\caption{Integrals of splitting functions $J^{(n)}(x)$ and 
$P^{(n)}_{\Delta}$.}
\begin{tabular}[]{|l|r|r|r|r|r|} \hline
& 
\multicolumn{1}{c|}{${\mathcal O}(\alpha)$}   & 
\multicolumn{1}{c|}{${\mathcal O}(\alpha^2)$} & 
\multicolumn{1}{c|}{${\mathcal O}(\alpha^3)$} & 
\multicolumn{1}{c|}{${\mathcal O}(\alpha^4)$} & 
\multicolumn{1}{c|}{${\mathcal O}(\alpha^5)$} \\ \hline
\multicolumn{1}{|c|}{$x$} &
\multicolumn{5}{|c|}{$J^{(n)}(x)$} \\ \hline 
0.01 &  $-$0.0101 &$-$0.0664 &    $-$0.1854 &  $-$0.1409 &         0.5232 \\
0.1  &  $-$0.1057 &$-$0.4529 &    $-$0.4487 &     1.6579 &         2.7010 \\
0.5  &  $-$0.7613 &$-$1.6997 &       5.7829 &     7.5130 &     $-$99.1465 \\
0.9  &  $-$3.3002 &   5.4338 &      31.0395 &$-$376.7425 &      2102.5492 \\
0.99 &  $-$7.7303 &  53.2969 &  $-$293.5939 &   708.7696 &     11810.0253 \\
0.999& $-$12.3175 & 145.1533 & $-$1606.9698 & 16122.5016 & $-$133803.0981 \\
\hline
\multicolumn{1}{|c|}{$\Delta$} &
\multicolumn{5}{|c|}{$P^{(n)}_{\Delta}$} \\ \hline 
0.001& $-$12.3155 & 145.0921 & $-$1605.5843 & 16095.0956 & $-$133303.9430 \\
\hline
\end{tabular}
\end{table}

\section*{Appendix A}

\setcounter{equation}{0}
\renewcommand{\theequation}{A.\arabic{equation}}

The Riemann $\zeta$--functions are defined as usual:
\begin{eqnarray} 
\zeta(n) &=& \sum\limits_{k=1}^{\infty}\frac{1}{k^n}, \qquad
\zeta(2) = \frac{\pi^2}{6}\, , \qquad 
\zeta(3) \approx 1.20205690315959, \nonumber \\
\zeta(4) &=& \frac{\pi^4}{90}\, , \qquad
\zeta(5) \approx 1.03692775514337\; .
\end{eqnarray}

Here we define the polilogarithms, which enter into our formulae.
We follow the notations of Ref.~\cite{KMR,DD}.
The general Nielsen's polilogarithm is
\begin{equation} 
\mathrm{S}_{n,m}(z) = \frac{(-1)^{n+m-1}}{(n-1)!m!}
\int\limits_{0}^{1}\frac{\dd x}{x}\ln^{n-1}(x)\ln^m(1-xz),
\end{equation}
in particular
\begin{eqnarray} 
\Li(z) &=& \mathrm{S}_{1,1}(z) = - \int\limits_{0}^{1}
\frac{\dd x}{x}\ln(1-xz), \qquad
\mathrm{S}_{1,2}(z) = \frac{1}{2}\int\limits_{0}^{1}\frac{\dd x}{x}
\ln^2(1-xz), 
\nonumber \\
\mathrm{Li}_3(z) &=& \mathrm{S}_{2,1}(z) 
= \int\limits_{0}^{1}\frac{\dd x}{x}\ln(x)\ln(1-xz) 
= \int\limits_{0}^{1}\frac{\dd x}{x}\Li(x),
\nonumber \\
\mathrm{Li}_4(z) &=& \mathrm{S}_{3,1}(z) 
= - \frac{1}{2} \int\limits_{0}^{1}\frac{\dd x}{x}\ln^2(x)\ln(1-xz),
\qquad
\mathrm{S}_{1,3}(z) 
= - \frac{1}{6} \int\limits_{0}^{1}\frac{\dd x}{x}\ln^3(1-xz),
\nonumber \\
\mathrm{S}_{2,2}(z) &=& - \frac{1}{2} \int\limits_{0}^{1}
\frac{\dd x}{x}\ln(x)\ln^2(1-xz).
\end{eqnarray} 

In order to get expansion~(\ref{expand}) it is convenient to use
the following representation for $\Gamma$-function:
\begin{eqnarray}
\frac{1}{\Gamma(z)} = z{\mathrm{e}}^{Cz}\prod_{n=1}^{\infty}
\biggl( 1+\frac{z}{n} \biggr){\mathrm{e}}^{-z/n}.
\end{eqnarray}

\end{document}